\documentclass[american,reprint,aps,pra,superscriptaddress,showpacs]{revtex4-1}
\usepackage[T1]{fontenc}
\usepackage[latin9]{inputenc}
\usepackage{babel}
\usepackage{amsthm}
\usepackage{amsmath}
\usepackage{amssymb}
\usepackage[unicode=true,pdfusetitle,
 bookmarks=true,bookmarksnumbered=false,bookmarksopen=false,
 breaklinks=false,pdfborder={0 0 0},backref=false,colorlinks=false]
 {hyperref}
\hypersetup{
 colorlinks,linkcolor=myurlcolor,citecolor=myurlcolor,urlcolor=myurlcolor}

\makeatletter
\@ifundefined{textcolor}{}
{%
 \definecolor{BLACK}{gray}{0}
 \definecolor{WHITE}{gray}{1}
 \definecolor{RED}{rgb}{1,0,0}
 \definecolor{GREEN}{rgb}{0,1,0}
 \definecolor{BLUE}{rgb}{0,0,1}
 \definecolor{CYAN}{cmyk}{1,0,0,0}
 \definecolor{MAGENTA}{cmyk}{0,1,0,0}
 \definecolor{YELLOW}{cmyk}{0,0,1,0}
 }
\theoremstyle{plain}

\ifx\proof\undefined
\newenvironment{proof}[1][\protect\proofname]{\par
\normalfont\topsep6\p@\@plus6\p@\relax
\trivlist
\itemindent\parindent
\item[\hskip\labelsep
\scshape
#1]\ignorespaces
}{%
\endtrivlist\@endpefalse
}
\providecommand{\proofname}{Proof}
\fi

\usepackage{times}
\usepackage{txfonts}
\usepackage{braket}
\usepackage{colortbl}
\definecolor{myurlcolor}{rgb}{0,0,0.7}

\makeatother

\providecommand{\theoremname}{Theorem}

\begin{document}
\newcommand{\beq}{\begin{equation}}
\newcommand{\eeq}{\end{equation}}
\bibliographystyle{apsrev}

\title{Duality of Quantum Coherence and Path Distinguishability}
\author{Manabendra Nath Bera}
\email{manabbera@hri.res.in}
\affiliation{Quantum Information and Computation Group, Harish-Chandra Research Institute, Allahabad-211019, India.}

\author{Tabish Qureshi}
\email{tabish@ctp-jamia.res.in}
\affiliation{Centre for Theoretical Physics, Jamia Millia Islamia, New Delhi-110025, India.}
\author{Mohd Asad Siddiqui}
\email{asad@ctp-jamia.res.in}
\affiliation{Centre for Theoretical Physics, Jamia Millia Islamia, New Delhi-110025, India.}

\author{Arun Kumar Pati}
\email{akpati@hri.res.in}
\affiliation{Quantum Information and Computation Group, Harish-Chandra Research Institute, Allahabad-211019, India.}


\begin{abstract}
We derive a generalized wave-particle duality relation for arbitrary multi-path quantum interference phenomena. Beyond the 
conventional notion of the wave nature of a quantum system, i.e., the interference fringe visibility, we introduce a novel quantifier as the normalized quantum coherence, recently defined in the framework of quantum information theory. To witness the particle nature, we quantify the path distinguishability or the which-path information
based on 
unambiguous quantum state discrimination. Then, the Bohr complementarity principle, for multi-path quantum interference, can be stated as a duality relation between the quantum coherence and the path distinguishability.   
For two-path interference, the quantum coherence is identical to the interference fringe visibility, and the relation reduces to the well-know complementarity relation. The new duality relation continues to hold in the case where mixedness is introduced due to possible decoherence effects.


\end{abstract}
\maketitle

\section{Introduction}

One of the famous, yet intriguing feature of quantum mechanics is the wave-particle duality. This is often captioned by Bohr's complementarity principle. It states  that the wave aspect and the particle aspect are
complementary in nature, in the sense that if an experiment clearly reveals the wave
nature, it will completely hide the particle aspect and vice-versa \cite{bohr}.
The complementarity principle has been a subject of debate since the time
of its inception when Einstein proposed his famous recoiling slit experiment
(see e.g. \cite{tqeinstein}).

Since then, attempts have been made to give a quantitative meaning to the
complementarity principle, in the context of interference experiments
\cite{wootters, greenberger, englert}. The idea is to investigate how much
of each aspect, wave or particle, can be seen at the same time. 
In two-path interference experiment,
either in a two-slit experiment or in a two-path Mach-Zehnder interferometer,
the principle of complementarity is quantitatively represented by the
 Englert-Greenberger-Yasin (EGY) relation
\begin{equation}
{\mathcal V}^2 + {\mathcal D}^2 \le 1,
\label{egy}
\end{equation}
where ${\mathcal V}$ is the visibility of the interference pattern and  ${\mathcal D}$ is a measure of  path distinguishability or which-path information.
For two-path interference, the quantum system (quanton) may arrive at the detector along two different paths. If the experimenter determines which path the quanton has traveled through without ambiguity (i.e., $\mathcal{D}=1$), then no interference fringes will be seen (i.e., $\mathcal{V}=0$). On the other hand, a non-zero ambiguity in the which-path information (i.e., $\mathcal{D}\neq 1$), of the experimenter, may retain a non-vanishing fringe visibility (i.e., $\mathcal{V} \neq 0$).    
Thus, the knowledge of which-path information or the path distinguishability limits the interference visibility ${\mathcal V}$ in an interference experiment, according to the above complementarity relation. This relation has been demonstrated experimentally with atoms \cite{Durr98}, nuclear magnetic resonance \cite{Peng03, Peng05}, faint lasers \cite{Schwindt99}, and also with single photons   \cite{Jacques08}. Further, the complementarity relation has been extended to the more general case of an asymmetric interferometer where only a single output port is considered and this duality holds \cite{Li12}. Recently, the duality relation has also been investigated in the presence of non-locality \cite{Peruzzo12} and quantum entanglement \cite{Jakob03}.           


Intuitively, the complementarity relation, between the wave and the particle nature of the quantons, is expected to hold in multi-path or multi-slit experiments too. Several attempts have been made to quantitatively formulate
the complementarity principle in multi-path experiments \cite{jaeger, durr, bimonte, englertmb}. However, a derivation of a loophole free generalized complementarity relation, for multi-slit quantum interference experiment, is still demanding.
The underlying problem, for that, is the absence of 
exact analytical forms of interference fringe visibility and path distinguishablity, which are strictly complementary to each other, for an $n$-path quantum interferometer. One may seek to resolve this difficulty by relaxing the conventional signatures of particle and wave natures of the quantons, such as which-path information and fringe visibility.    
While the fringe visibility being the most used signature of wave nature of the quantum particles, it is certainly not the only one to capture the essence. For example, recently, the wave-particle duality in two-slit experiments has been shown to be equivalent to the entropic uncertainty relation \cite{coles}, and the wave and particle natures are represented with entropic quantifiers.

In this paper we derive a generalized complementarity relation from such an alternative perspective. In our approach, we quantify the wave nature in terms quantum coherence, which has been proposed recently in the context of quantum information theory \cite{coherence}. On the other hand, the particle nature connected to the which-path information or the path distinguishability, is quantified by the upper bound of the  success probability in the unambiguous quantum state discrimination (UQSD) \cite{Helstrom76, Bae15, Pati05}. Remarkably, the quantum coherence and the path distinguishability are truly complementary in nature. That means, an increase in quantum coherence is always associated with a decrease in path distinguishability and vice versa.
With the proposed quantifiers, we derive a generalized complementarity relation for arbitrary $n$-slit scenario for both pure and mixed quanton states. We show that, the sum of (normalized) quantum coherence and path distinguishability in the complementarity relation, exactly, equals to one for every pure states and upper bounded by the same in the case of mixed states. Our duality relation then gives a 
justification to the measure of quantum coherence as it truly brings out the wave nature of the quanton at its heart.


The paper is organized as follows. In section \ref{sec:coh}, we introduce the quantifiers of quantum coherence and unambiguous quantum state discrimination which quantitatively capture the wave nature and  particle nature of the quantum system, respectively. The new duality relation, for pure quanton and detector state is derived in \ref{sec:pureCR}. Then we generalize the duality relation for mixed states of arbitrary dimension. 
Finally, we conclude in section \ref{sec:concl}.

\section{Quantum Coherence and UQSD \label{sec:coh}}
\subsection{Quantum coherence}
Coherence is a fundamental feature of quantum physics which signifies the possible superposition between the orthogonal quantum states. Again, it is largely believed that, the quantum superposition is the manifestation of wave nature of quantum particles.  Thus, the quantum coherence has a strong correspondence with the wave nature of a quantum particle. Though a rigorous study of coherence has been carried out in quantum optics in terms of quasi-probabilities, a generalized quantification of quantum  coherence was absent until recently. In Ref. \cite{coherence}, Baumgratz {\em et al.}, proposed a reliable quantifier of quantum coherence from quantum information theoretic approach. The framework, to quantify coherence, is based on the characterization of the set of incoherent quantum states ($\mathcal{I}$).
For a given incoherent basis $\{ |i\rangle \}$, the incoherent states are defined as $\sigma^{\mathcal{I}}=\sum_i p_i |i\rangle \langle i| \in \mathcal{I}$, where $p_i$s are non-negative probabilities with $\sum_i p_i=1$. The 
incoherent operations are the completely positive trace preserving (CPTP) maps $\Lambda^{\mathcal{I}}$, those transform $\sigma^{\mathcal{I}}\rightarrow \Lambda^{\mathcal{I}}(\sigma^{\mathcal{I}}) \in \mathcal{I}$ to an incoherent state which is, again, diagonal in the incoherent basis. The maximally coherent state, of dimension $n$, is defined as $|\Psi\rangle = {1\over\sqrt{n}}\sum_{i=1}^n |i\rangle$ and the coherence of such state is used as reference to compare the coherence in the other states. For a given incoherent basis, the reliable quantifier of quantum coherence $\mathcal{C}(\rho)$, is a function of the state $\rho$, and should satisfy \cite{coherence}: (1) $\mathcal{C}(\rho)=0$ if and only if $\rho \in \mathcal{I}$, (2) $\mathcal{C}(\rho)$ is non-increasing under incoherent operations, i.e., $\mathcal{C}(\rho) \geqslant \mathcal{C}(\Lambda^{\mathcal{I}} \rho)$, (3) $\mathcal{C}(\rho)$ is non-increasing on an average under selective incoherent measurement, i.e., $\mathcal{C}(\rho)\geqslant \sum_m q_m \mathcal{C}(\rho_m)$, 
where $\rho_m=\frac{1}{q_m}K_m \rho K_m^\dag$,  \ $q_m=\mbox{Tr}(K_m \rho K_m^\dag)$, $K_m$'s are the Kraus operators and $K_m \mathcal{I} K_m^\dag \subseteq \mathcal{I}$, (4) $\mathcal{C}(\rho)$ is non-increasing under convex mixing of density matrices, i.e., $\mathcal{C}(\sum_k q_k \rho_k)\leqslant \sum_k q_k \mathcal{C}(\rho_k)$. The functions of density matrix those satisfy these properties are relative entropy of coherence, $l_1$-norm of coherence \cite{coherence} and skew information of coherence \cite{Girolami}. In this work, we shall use $l_1$-norm of coherence to quantify the wave nature of the quantons, defined as
\begin{equation}
 \mathcal{C}(\rho)=\mbox{min}_{\sigma^\mathcal{I} \in \mathcal{I}}\parallel \rho-\sigma^\mathcal{I} \parallel_{l_1},
\end{equation}
where $\parallel A \parallel_{l_1}=\sum_{ij}|A_{ij}|$ and the minimization is carried out over the set of all incoherent states $\sigma^\mathcal{I}$. It can easily be seen that the minimization is achieved for the $\sigma^\mathcal{I}=\sum_i \langle i | \rho | i \rangle |i \rangle \langle i|$ and then the coherence quantitatively becomes the sum of the absolute values of the off-diagonal elements of the
density matrix of a system, i.e., 
\begin{equation}
 \mathcal{C}(\rho)=\sum_{i\neq j} |\rho_{ij}|
\end{equation}
with $\rho_{ij}=\langle i | \rho | j \rangle$.
Using this measure, we define normalized quantum coherence
as
\begin{equation}
{\mathcal C}(\rho) = {1\over n-1}\sum_{i\neq j} |\rho_{ij}| ,
\label{coherence}
\end{equation}
where $n$ is the dimensionality of the Hilbert space. Hereafter we call normalized coherence as ``coherence'', for simplicity. In what follows, we shall show that the coherence captures the wave nature of a quanton in a multi-path quantum interference scenario.
\subsection{Unambiguous Quantum State Discrimination \label{sec:uqsd}}
Quantum state discrimination has various important applications in quantum information theory          
\cite{Helstrom76, Bae15, Pati05}. In the quantum state discrimination, the task is to find out which state an experimenter has in her possession to the best of her ability \cite{Helstrom76}. In quantum mechanics the existence of non-orthogonal states adds further difficulties to the problem, in addition to the statistical mixing of the quantum states. The commonly used strategies to discriminate non-orthogonal states can be divided in two broader classes, those are the ambiguous \cite{Helstrom76} and the unambiguous \cite{uqsd} quantum state discrimination. In ambiguous state discrimination, one always has an answer but with a probability of being wrong and the task is to minimize this probability. On the other hand, in unambiguous state discrimination, one guarantees to never be wrong but sometime may have nonanswer, that is to say one does not know. In this case the task is to minimize the probability of a nonanswer. The unambiguous state discrimination is particularly interesting for the cases where the states analyzed are mixed \cite{Rudolph03}. However, in what follows, we shall stick to UQSD for pure states as it suffices for our analysis.



UQSD was first formulated for unambiguously discriminating between two
non-orthogonal states \cite{uqsd}. Consider a state which could be either
$|d_1\rangle$ or $|d_2\rangle$, with equal probability. The probability with
which one can tell {\em for sure} which of the two states is the given one,
is bounded by \cite{uqsd}
\begin{equation}
P_2 \le 1 - |\langle d_1|d_2\rangle|.
\label{p2}
\end{equation}
One can also precisely specify the condition in which the success probability is
\begin{equation}
P_2 = 1 - |\langle d_1|d_2\rangle|,
\label{idp}
\end{equation}
which is the so-called IDP (Ivanovic-Dieks-Peres) limit. The two states,
$|d_1\rangle$ and $|d_2\rangle$, cannot be unambiguously discriminated
with a probability larger than the IDP limit, even in principle. 
This shows that the IDP limit sets a fundamental limit to distinguish
two non-orthogonal states $|d_1\rangle$ and $|d_2\rangle$.


UQSD was later generalized to $n$ non-orthogonal states \cite{zhang}.
Consider a quantum state prepared in one of the $n$ states $|d_1 \rangle, ..., |d_n \rangle$ in an $n$-dimensional Hilbert space with corresponding probabilities $p_1,...,p_n$. The states are in general non-orthogonal. To find out which of the $n$ states, the given state is, one needs to perform one or more quantum measurements. The upper bound for the success probability, of unambiguous discrimination among the $n$ quantum states, is given by \cite{zhang}
\begin{equation}
P_n \le 1 - {1\over n-1}\sum_{i\neq j} \sqrt{p_ip_j} |\langle d_i|d_j\rangle|.
\label{pn}
\end{equation}
Clearly, for orthogonal quantum states, there exists a quantum measurement strategy for which the success probability of UQSD reaches maximum to 1.
For non-orthogonal states, there may exist a measurement strategy for which the 
success probability of UQSD is
\begin{equation}
P_n = 1 - {1\over n-1}\sum_{i\neq j} \sqrt{p_ip_j} |\langle d_i|d_j\rangle|.
\label{pnmax}
\end{equation}
However, since the above does not represent the optimal success probability,
this upper bound may, in general, not be achievable.
In the next section, we shall use this upper bound of the success probability for UQSD as a measure of which-path information and that in turn can witness the particle nature of the quanton.

\section{Complementarity  of Coherence and Path Distinguishability \label{sec:pureCR}}
\subsection{Duality relation for pure quanton and detector states}
Let us now consider the case of a $n$-slit quantum interference with pure
quantons. In $n$-slit interference, if $|\psi_i\rangle$ is the possible
state of the quanton if it takes the $i$'th slit or $i$'th path, then state of
the quanton, after crossing the slit, can be described solely in terms
of $|\psi_i\rangle$, and it may be treated as a basis state. Thus the
state of the quanton can be represented in terms of $n$ basis states 
$\{|\psi_1\rangle,|\psi_2\rangle,\dots,|\psi_n\rangle \}$, where each of
the states represents each slit or path.
\begin{equation}
|\Psi \rangle = c_1|\psi_1\rangle + c_2|\psi_2\rangle  +
\dots + c_n|\psi_n\rangle ,
\label{ent}
\end{equation}
and $c_i$ is the amplitude of taking the i'th path.
In an interference experiment,
if one wants to know which of the $n$ slits the quanton passes through,
or which of the $n$ paths does the quanton take, one needs to perform
a quantum measurement.
In quantum measurements, according to von Neumann, the first process is
to let a detector interacts with a quanton and gets entangled with it
\cite{neumann}. Then, the quantum measurements may be performed on the
detector state to infer about the properties of the quanton.
In general, the controlled unitary operations are used to correlate the
quanton and detector in an interference experiment. For a quanton
state $| \Psi\rangle$ and controlled unitary
interaction $U(|\psi_i\rangle |0_d\rangle) \mapsto |\psi_i\rangle |d_i\rangle$,
where $|0_d\rangle$ is the initial detector state, the combined
quanton-detector state becomes
\begin{equation}
|\Psi\rangle = c_1|\psi_1\rangle \otimes  |d_1\rangle + c_2|\psi_2\rangle  \otimes |d_2\rangle +
\dots , c_n|\psi_n\rangle \otimes  |d_n\rangle,
\label{ent}
\end{equation}
where $|d_i\rangle$ is the state of the which-path detector if the quanton
went through the $i$'th path, and $\sum_{i=1}^n|c_i|^2=1$.
For simplicity, we consider the detector states
$\{|d_i\rangle\}$ to be normalized, but not necessarily orthogonal.
Now, if one tries to acquire knowledge about which path the quanton took, it shall reduce the coherence of the quanton. 
The left out coherence present in the quanton will correspond to the coherence of the reduced density matrix of the quanton, where the latter is given by
\begin{equation}
\rho_s = \sum_{i=1}^n\sum_{j=1}^nc_ic_j^* \ \langle d_j|d_i\rangle \ |\psi_i\rangle\langle\psi_j|,
\end{equation}
after tracing over the detector states.
%
For a given interferometric set-up the set $\{ |\psi_i\rangle\}$ forms the incoherent bases. The  coherence can now be calculated for the particle using the reduced 
density matrix given above, as
\begin{eqnarray}
{\mathcal C} &=& {1\over n-1}\sum_{i\neq j}|\langle\psi_i|\rho_s|\psi_j\rangle|
\nonumber\\
&=& {1\over n-1}\sum_{i\neq j} |c_i||c_j||\langle d_j|d_i\rangle|.
\label{Cn}
\end{eqnarray}
It is interesting to note that if the detector states $\{|d_i\rangle \}$ form a mutually orthogonal basis, the reduced system states becomes diagonal in the incoherent basis and hence, has vanishing coherence. This implies that, in this situation, the 
wave nature of the quanton cannot be seen. However, the situation will be different if the detector states are not mutually orthogonal to each other and in the reduced density matrix, the off-diagonal elements not necessarily be vanishing. Thus, the wave aspect of the quanton will acquire non-zero value as the quantum coherence is non-vanishing. But, the coherence of the quanton will be certainly reduced than that of the original state before the measurement interaction was turned on.

Now, let us focus on the problem of path distinguishability or which-path information which is attributed to the particle nature of the quantons.
Since, through the controlled unitary interaction, each of the path is marked with a detector state $| d_i \rangle$, the path distinguishability is equivalent to discriminating the detector states. In other words, if the quanton passes through the $i$'th path, the resulting the detector states becomes $|d_i\rangle$ with the probability $|c_i|^2$. Now distinguishing all these detector states $\{|d_i\rangle \}$ with the corresponding probabilities $\{|c_i|^2 \}$, is equivalent to distinguishing the paths the quanton chooses in the interferometric set-up. If the detector states $\{|d_i\rangle \}$ are mutually orthogonal, then the states can be distinguished with unit probability. In this case we will know which path the system has taken with certainty. However, the interesting case appears when $\{|d_i\rangle \}$ are not mutually orthogonal and in that case we have partial knowledge about which path information.
In general, the best strategy to distinguish between non-orthogonal states
is unambiguous quantum state discrimination (UQSD)
\cite{uqsd, jaeger2, bergou}. 
In UQSD, the success probability with which non-orthogonal pure states can
be {\em unambiguously} distinguished depends on the measurement strategies
employed. One would like to know which strategy yields the maximum
success probability. The optimal success probability of unambigiously
distinguishing between $n$ non-orthogonal states is not known. However,
the success probability in the UQSD between the detector states $\{ | d_i \rangle \}$ with corresponding probabilities $\{|c_i|^2 \}$, is {\em bounded by} \cite{zhang}
\begin{equation}
P_n \le 1 - {1\over n-1}\sum_{i\neq j} |c_i||c_j| |\langle d_i|d_j\rangle|.
\label{pn}
\end{equation}
Note that the probabilities $|c_i|^2$ are decided by the initial superposition in the quanton state.
The upper bound
given by (\ref{pn}) may not be achievable in practice in many situations.
However, it can still serve as a measure of the degree to which the $n$ states
are distinguishable, namely the $n$ states cannot be distinguished with a
probability larger than that given by the upper bound of (\ref{pn}).
Here we {\em define} the path-distinguishability ${\mathcal D}_Q$ as the upper bound of the success
probability with which the $n$ paths of the particle can be distinguished
without any error. The subscript $Q$ is added to differentiate it from the
${\mathcal D}$ used in \cite{englert} and elsewhere in the literature.
This is just the upper bound of the success probability with which the states $\{|d_i\rangle\}$
can be unambiguously discriminated, which is the saturation limit of
(\ref{pn}). Thus, the path-distinguishability, for $n$-path interference, can
be defined as
\begin{equation}
{\mathcal D}_Q := 1 - {1\over n-1}\sum_{i\neq j} |c_i||c_j| |\langle d_i|d_j\rangle|.
\label{D}
\end{equation}
The path-distinguishability can take values between 0 and 1. For all mutually orthogonal detector states $\{|d_i\rangle\}$, one has ${\mathcal D}_Q=1$.

Now with the quantum coherence left in the reduced quanton state, in Eq. (\ref{Cn}) and the path-distinguishability, defined in Eq. (\ref{D}), 
we get a general $n$-slit duality relation, for arbitrary pure quantons, as
\begin{equation}
{\mathcal C} + {\mathcal D}_Q = 1 .
\label{duality}
\end{equation}
Using the ${\mathcal C}$ and ${\mathcal D}_Q$ as the quantifiers of the wave and the particle nature of the quanton, respectively, the 
Eqn. (\ref{duality}) puts a bound on how much of wave nature and particle nature
a system can display at the same time. This can be treated as a
quantitative statement of Bohr's principle, for $n$-path interference, using the measures of quantum coherence and path distinguishability. Note, the ${\mathcal C}$ and ${\mathcal D}_Q$ are truly complementary in nature where an increase in one results in a decrease in the other.

%

\subsubsection{Two-slit interference}

In a two-path interference with equally probable paths we have
$|c_1|=|c_2|=\frac{1}{\sqrt{2}}$, and hence the path-distinguishability becomes
${\mathcal D}_Q=1-|\langle d_1|d_2\rangle|$ which, interestingly, is just the
IDP limit. The coherence in (\ref{Cn}) reduces to
\begin{equation}
{\mathcal C} = |\langle d_1|d_2\rangle|.
\label{C2}
\end{equation}
However, for a double-slit experiment, the interference visibility,
defined as ${\mathcal V} \equiv {I_{max} - I_{min} \over I_{max} + I_{min} }$,
where $I_{max}$ and $I_{min}$ represent the maximum and minimum intensity
in neighboring fringes, respectively, is just 
${\mathcal V} = |\langle d_1|d_2\rangle|$ \cite{tqeinstein}. Therefore, the fringe
visibility is just equal to the coherence ${\mathcal C}$. The duality
relation (\ref{duality}) now becomes
\begin{equation}
 {\mathcal V} + {\mathcal D}_Q = 1 .
\end{equation}
In practical scenario there may be other factors which reduce the visibility of
fringes in interference experiments. So in the general case, the above relation will be an 
{\em inequality}, saturating to equality.
This inequality has been derived before \cite{3slit}, and is completely
equivalent to the EGY duality relation (\ref{egy}) with the 
recognition that ${\mathcal D_Q}  = 1 -  \sqrt{ 1 - {\mathcal D}^2}$
\cite{3slit}.
Therefore, for the two-slit case, the new duality relation (\ref{duality})
reduces to the EGY relation (\ref{egy}).

\subsubsection{Three-slit interference}

For the three-slit interference, the path-distinguishability becomes
${\mathcal D}_Q = 1 - (|c_1||c_2| |\langle d_1|d_2\rangle | +
|c_2||c_3| |\langle d_2|d_3\rangle | + |c_1||c_3| |\langle d_1|d_3\rangle |)$,
and the coherence reduces to 
\begin{equation}
{\cal C} = |c_1||c_2||\langle d_1|d_2\rangle|
+|c_2||c_3| |\langle d_2|d_3\rangle |+|c_1||c_3| |\langle d_1|d_3\rangle|.
\label{C3}
\end{equation}
%
%
%
This coherence can be shown to be related to the ideal interference
visibility by the relation \cite{3slit}
\begin{equation}
{\mathcal C} = {2{\mathcal V}\over 3 - {\mathcal V}}.
\end{equation}
The duality  relation reduces to
\begin{equation}
{\mathcal D}_Q + {2{\mathcal V}\over 3 - {\mathcal V}} = 1,
\end{equation}
which is exactly the same as the duality relation derived for the 
three-slit interference \cite{3slit}. One should of course realize that
in a non-ideal situation, the fringe visibility will be reduced, and the the above relation will be an {\em inequality}.
Thus, for the 3-slit case, the new duality relation reduces to an earlier
relation derived independently.

\subsection{Generalization of duality relation for mixed states \label{sec:mixCR}}


We now extend the preceding analysis to the situations where a certain amount
of mixedness is introduced in density matrix of the quanton state, say $\rho=\sum_{ij} \rho_{ij}|\psi_i \rangle \langle \psi_j|$.
This may happen if the quantum system is exposed to environment. Interference experiments have been
carried out with large molecules, where the interaction with the environment,
although minimized, is not fully avoided \cite{c60, c70}. In a scenario where the initial detector state is chosen to be pure, the combined density matrix of the quanton and the path-detector after the controlled unitary (the measurement interaction), may be written as
\begin{equation}
\rho_{sd} = \sum_{i=1}^n\sum_{j=1}^n \rho_{ij}|\psi_i\rangle\langle\psi_j| \otimes |d_i\rangle\langle d_j|,
\label{rhom}
\end{equation}
where the quanton state becomes entangled with the detector. Note that the combined state is also mixed since the initial quanton state is mixed. As in the pure state scenario, the wave nature in the post-interaction quanton state can be quantified with  the quantum coherence present in the reduced quanton state.   
By tracing out over the path-detector states in (\ref{rhom}), one gets
the reduced density matrix for the quanton
\begin{equation}
\rho_{s} = \sum_{i=1}^n\sum_{j=1}^n \rho_{ij}|\psi_i\rangle\langle\psi_j|
\langle d_j|d_i\rangle .
\label{rhomr}
\end{equation}
The coherence can now be calculated as
\begin{eqnarray}
{\mathcal C} &=& {1\over n-1}\sum_{i\neq j}|\langle\psi_i|\rho_{s}|\psi_j\rangle|
\nonumber\\
&=& {1\over n-1}\sum_{i\neq j} |\rho_{ij}||\langle d_j|d_i\rangle|.
\label{Cm}
\end{eqnarray}
Note that, before the interaction between quanton and detector, the coherence is ${\mathcal C}={1\over n-1}\sum_{i\neq j} |\rho_{ij}|$, which is reduced after the interaction due to the factors $|\langle d_j|d_i\rangle| \leqslant 1$.

Again the particle nature of the quantons can be expressed, quantitatively, by the path-distinguishability or which-path information. That is nothing but how well an experimenter can distinguish between the detector states $\{|d_i\rangle \}$ with the corresponding probabilities $\{ \rho_{ii} \}$, where the probabilities are determined by the initial quanton state. 
One may carry out UQSD on the state $|d_i\rangle$ as before. Since a state
$|d_i\rangle$ appears with a probability $\rho_{ii}$, the path distinguishability for
$n$-path interference which is the upper bound of success probability in UQSD, in the mixed case, can be written as
\begin{eqnarray}
{\mathcal D}_Q  &=& 1 - {1\over n-1}\sum_{i\neq j} \sqrt{\rho_{ii}\rho_{jj}} |\langle d_i|d_j\rangle|.
\label{Dm}
\end{eqnarray}

For a given quanton state, the path distinguishability and the quantum coherence of the quanton depend on the choice of initial state of the which-path detector and the measurement interaction.  Now from  (\ref{Cm}) and (\ref{Dm}), we get
\begin{equation}
{\mathcal C} + {\mathcal D}_Q + {1\over n-1}\sum_{i\neq j}
(\sqrt{\rho_{ii}\rho_{jj}} - |\rho_{ij}|)|\langle d_j|d_i\rangle| = 1 .
\label{c+ds}
\end{equation}
Since every principal 2x2 sub matrix of (\ref{rhom}) is positive
semi-definite \cite{horn}, we have
\begin{equation}
\sqrt{\rho_{ii}\rho_{jj}} - |\rho_{ij}| \ge 0,
\label{eq:se}
\end{equation}
for arbitrary $i$ and $j$. This in turn implies that
${1\over n-1}\sum_{i\neq j}
(\sqrt{\rho_{ii}\rho_{jj}} - |\rho_{ij}|)|\langle d_j|d_i\rangle| \ge 0$. The equality holds only for the initial pure quantum systems. 
Thus, (\ref{c+ds}) leads to the following duality relation between  the
quantum coherence and the path distinguishability which are the quantifiers of wave nature and particle nature of a quanton respectively, given by
\begin{equation}
{\mathcal C} + {\mathcal D}_Q \le 1.
\label{c+d}
\end{equation}
The above is the generalized version of (\ref{duality}) and applicable for any mixed quanton state. 
One can easily see that, in an experiment where the mutual overlap between the detector states are simultaneously increased or decreased, these two quantities become truly complementary in nature. There, an increase in the path distinguishability
inevitably reduces the quantum coherence in the quanton state, and vice versa.


Our analysis can further be extended in the case where one has the initial detector in a mixed state, as well. For the controlled-unitary operation as the measurement interaction $U=\sum_{i}|\psi_i\rangle \langle \psi_i| \otimes U_i$ and initial detector state $\rho_d$, the combined quanton-detector state after the interaction becomes
\begin{equation}
 \rho_{sd}=\sum_{ij}\rho_{ij}|\psi_i\rangle \langle \psi_j|\otimes U_i \rho_d U_i^\dag.
\end{equation}
If the measurement interaction leads to mutually orthogonal detector states, then $\mbox{Tr}\left( U_i\rho_d U_i^\dag U_j\rho_d U_j^\dag\right)=0$ for $\forall \ i\neq j$. In such situation, one may easily see that $\mbox{Tr}\left( U_i \rho_d U_j^\dag \right)=0$, for $\forall \ i\neq j$, holds. The reduced quanton state after the interaction can be written as
\begin{equation}
 \rho_s^\prime = \sum_{i,j}^n \rho_{ij} \ \mbox{Tr}\left( U_i \rho_d U_j^\dag \right) \ |\psi_i\rangle \langle \psi_j|. 
\end{equation}
Clearly, for a ``good'' measurement interaction for which $\mbox{Tr}\left( U_i \rho_d U_j^\dag \right)=0$ for $\forall \ i\neq j$, the quanton state reduces to an incoherent state and thus the wave nature becomes absent.
On the other hand if the  $\mbox{Tr}\left( U_i \rho_d U_j^\dag \right) \neq0$, there remains non-vanishing quantum coherence in the reduced quanton state, as 
\begin{equation}
 \mathcal{C}^\prime =\frac{1}{n-1} \sum_{i\neq j} \left| \ \rho_{ij}  \ \mbox{Tr}\left( U_i \rho_d U_j^\dag \right) \right|.
\end{equation}
For an initial detector state, let $\rho_d=\sum_k r_k |d_k \rangle \langle d_k|$ be the spectral decomposition. Then, we have $\left| \ \rho_{ij} \ \mbox{Tr}\left( U_i \rho_d U_j^\dag \right)\right|=\left|\rho_{ij} \ \sum_k r_k \langle d_{ki}  | d_{kj} \rangle \right| \leqslant \sum_k r_k |\langle d_{ki}  | d_{kj} \rangle | |\rho_{ij} |$, where we denote $U_i |d_k \rangle=|d_{ki} \rangle$. Thus we are lead to
\begin{equation}
 \mathcal{C}^\prime \leqslant \frac{1}{n-1} \sum_k r_k \sum_{i\neq j} |\rho_{ij}|  |\langle d_{ki}  | d_{kj} \rangle |.
 \label{eq:mnc}
\end{equation}
Now for a given initial detector state $|d_k\rangle$, the success probability of UQSD for the ensemble $\{\rho_{ii}, \ |d_{ki}\rangle \}$ is bounded by $\mathcal{D}_Q^k=1-\frac{1}{n-1}\sum_{i\neq j} \sqrt{\rho_{ii}\rho_{jj}} |\langle d_{ki}  | d_{kj} \rangle |$, which represents the path distinguishability. Then, for the initial detector state $\rho_d=\sum_k r_k |d_k \rangle \langle d_k|$, the path distinguishability can be computed by averaging over the individual $\mathcal{D}_Q^k$s, i.e., 
\begin{align}
 \mathcal{D}_Q^\prime &= \sum_k r_k \mathcal{D}_Q^k \nonumber \\
                      &=1-\frac{1}{n-1} \sum_k r_k \sum_{i\neq j} \sqrt{\rho_{ii}\rho_{jj}} |\langle d_{ki}  | d_{kj} \rangle |.
                      \label{eq:mpd}
\end{align}
Now, using  Eqs. (\ref{eq:se}), (\ref{eq:mnc}) and (\ref{eq:mpd}),
we arrive at the most general duality relation
\begin{equation}
 \mathcal{C}^\prime + \mathcal{D}_Q^\prime \leqslant 1.
 \label{eq:genDuality}
\end{equation}
Thus, Eqs. (\ref{duality}), (\ref{c+d}) and (\ref{eq:genDuality}) constitute the central results in our paper.
It is important to note that, for initial mixed quanton and mixed detector states, 
a generalized complementarity relation holds, where the particle and wave natures are quantified via path distinguishability based on UQSD and the quantum coherence, respectively.

%

\section{Conclusions \label{sec:concl}}
In this paper, we have introduced a generalized duality relation for arbitrary dimensional multi-slit quantum interference experiments. To delineate the wave nature of the quanton, which passes through the interferometer, we define normalized quantum coherence based on the recently introduced quantifier of quantum coherence in the framework of quantum information theory \cite{coherence}. Since, both the interference and quantum coherence rely on quantum superposition of the quantum states, we claim that the proposed measure of (normalized) quantum coherence can be a quantifier of the wave-nature of the quanton, instead of traditional quantifier based on the interference fringe visibility. The particle nature of a quanton is associated with the which-path information acquired through the detection  process, i.e., which-path detection. In this work, we quantify the which-path information or path distinguishability by identifying it with the upper bound of success probability in unambiguous quantum state discrimination \cite{Helstrom76, Bae15, Pati05, zhang} of the detector states, after the path detectors are placed and the measurement interaction is turned on. 
Based on the normalized quantum coherence and the path distinguishability as the 
quantifiers the wave and particle natures of a quanton respectively, 
we derive a new duality relation, 
which is a quantitative statement of Bohr's 
principle of complementarity. For two-path and three-path  interference we 
have related quantum coherence to the fringe visibility and recovered the
corresponding known duality relations. Furthermore, we show that, in cases where
decoherence may introduce some mixedness in the density matrix of
the quanton as well as in the detector, the duality relation continues to hold. We hope that our results will have fundamental implications in understanding the quantum complementarity, in particular, the wave nature of a quantum system in terms of quantum coherence.

\section*{Acknowledgement}
The authors thank A. K. Rajagopal for many fruitful discussions. MNB gratefully acknowledges the support provided by the Centre for Theoretical Physics, Jamia Millia Islamia, New Delhi, during his visit, where this work was initiated. MAS thanks the University Grants Commission, India for financial support.

\end{document}